\documentclass[twocolumn,nofootinbib,amsmath,amssymb,aps,prd,balancelastpage,superscriptaddress]{revtex4-1}

\usepackage{xcolor}
\usepackage[active]{srcltx}
\usepackage{amsmath,amsfonts,amssymb,amsthm,amstext,amscd,eucal,srcltx}
\usepackage{epsfig,graphicx,bm}
\usepackage{epstopdf, epsf}
\usepackage{dcolumn}
\usepackage{comment}
\usepackage{hyperref}

\newcommand{\be}{\begin{equation}}
\newcommand{\ee}{\end{equation}}

\newcommand{\bse}{\begin{subequations}}
\newcommand{\ese}{\end{subequations}}
\newcommand{\bea}{\begin{eqnarray}}
\newcommand{\eea}{\end{eqnarray}}
\newcommand{\ba}{\begin{array}}
\newcommand{\ea}{\end{array}}
\newcommand{\bc}{\begin{center}}
\newcommand{\ec}{\end{center}}

\begin{document}
\preprint{IPM/P-2012/009}  
\vspace*{3mm}

\title{Strongest atomic physics bounds on Non-Commutative Quantum Gravity Models
}%



\author{{Kristian Piscicchia$^{b,c}$, Andrea Addazi$^{a,c}$, Antonino Marcian\`{o}$^{d,c}$, Massimiliano Bazzi$^c$, Michael Cargnelli$^{e,c}$,  Alberto Clozza$^c$, Luca De Paolis$^c$, Raffaele Del Grande$^{f,c}$, Carlo Guaraldo$^c$, Mihail Antoniu Iliescu$^c$, Matthias Laubenstein$^g$, Johann Marton$^{e,c}$, Marco Miliucci$^c$, Fabrizio Napolitano$^c$, Alessio Porcelli$^{e,c}$, Alessandro Scordo$^c$, Diana Laura Sirghi$^{c,h}$, Florin Sirghi$^{c,h}$, Oton Vazquez Doce$^c$, Johann Zmeskal$^{e,c}$ and Catalina Curceanu$^{c,h}$}\\ 
\vspace{0.5 cm}
{\it$^a$ Center for Theoretical Physics, College of Physics Science and Technology, Sichuan University, 610065 Chengdu, China}\\
{addazi@scu.edu.cn}\\
{\it$^b$ Centro Ricerche Enrico Fermi - Museo Storico della Fisica e Centro Studi e Ricerche “Enrico Fermi”, Roma, Italy, EU}\\
{\it$^c$ Laboratori Nazionali di Frascati INFN, Frascati (Rome), Italy, EU}\\
{\it$^d$ Center for Field Theory and Particle Physics \& Department of Physics\\
Fudan University, Shanghai, China} \\
{marciano@fudan.edu.cn} \\
{\it$^e$ Stefan Meyer Institute for subatomic physics, Austrian Academy of Science, Austria, EU}\\
{\it$^f$ Physik Department E62, Technische Universität München, 85748 Garching, Germany, EU}\\
{\it$^g$ Laboratori Nazionali del Gran Sasso INFN, Assergi (L'Aquila), Italy, EU}\\
{\it$^h$ IFIN-HH, Institutul National pentru Fizica si Inginerie Nucleara Horia Hulubei, Romania, EU}
}

\vspace{0.5 cm}

\begin{abstract}
\noindent
Investigations of possible violations of the Pauli Exclusion Principle represent critical tests of the microscopic space-time structure and properties. Space-time non-commutativity provides a class of universality for several Quantum Gravity models.  In this context the VIP-2 Lead experiment sets the strongest bounds, searching for Pauli Exclusion Principle violating atomic-transitions in lead, excluding the $\theta$-Poincar\'e Non Commutative Quantum Gravity models far above the Planck scale for non-vanishing $\theta_{\mu \nu}$ ``electric-like'' components, and up to $6.9 \cdot 10^{-2}$ Planck scales if $\theta_{0i} = 0$.
\end{abstract}

\maketitle

{\it Introduction ---} The Pauli Exclusion Principle (PEP) forbids two or more fermions to occupy the same quantum state. This simple but elegant principle is one of the main pillars of modern science, explaining processes in particle and nuclear physics, in astrophysics and biology.
%
%
As  proved by {\it W. Pauli} \cite{Pauli:1940zz}, PEP is a direct consequence of the Spin-Statistics theorem (SST) and arises from anti-commutation rules of fermionic spinor fields, in the construction of the Fock space of the theory. 

A fundamental assumption of SST is the Lorentz invariance, which strongly connects PEP to  the fate of the space-time symmetry and structure. Lorentz Symmetry may be dynamically broken at a very high energy scale $\Lambda$, without this implying a fundamental breakdown of the symmetry itself, generating non-renormalizable operators suppressed as inverse powers of $\Lambda$. Another possibility, present in several approaches to Quantum Gravity, is the {\it non-commutativity} of space-time coordinates close to the Planck scale, where Lorentz algebra turns out to be {\it deformed} at the very fundamental level. Space-time non-commutativity, as an extension of the uncertainty principle, is usually accredited to {\it W. Heisenberg} \cite{Heisenberg},  the idea being later elaborated by {\it H. Snyder} and {\it C.N. Yang} in Refs.~\cite{Snyder,Yang}. A symplectic-geometry approach \cite{SymGeo} unveils the deep relation intertwining space-time symmetries, spin-statistics  \cite{Arzano:2007gr} and the uncertainty principle \cite{SymGUP}, hence providing concrete path-ways for falsification.
 
Space-time non-commutativity is common to several Quantum Gravity frameworks, to which we refer as Non-Commutative Quantum Gravity models (NCQG). The connection of space-time non-commutativity with both String Theory (ST) \cite{Frohlich:1993es,Chamseddine1,Frohlich:1995mr,Connes:1997cr,Seiberg:1999vs} and Loop Quantum Gravity (LQG) \cite{AmelinoCamelia:2003xp,Freidel:2005me,Cianfrani:2016ogm,Amelino-Camelia:2016gfx, NClimLQG,BRAM, Brahma:2017yza} was extensively studied in literature. 

The two main classes of non-commutative space-time models embedding deformed Poincar\'e symmetries are characterized by $\kappa$-Poincar\'e \cite{Agostini:2006nc,13, AmelinoCamelia:2007uy, Arzano:2016egk} and $\theta$-Poincar\'e \cite{Addazi:2017bbg, AmelinoCamelia:2007wk,AmelinoCamelia:2007rn,Addazi:2018jmt,Addazi:2019ruk,Addazi:2018ioz} symmetries. 
Among these latter, there exists a sub-class of models which preserves the unitarity of the S-matrix in the Standard Model sector  \cite{Addazi:2017bbg, AG, Addazi:2018jmt}. 

From the experimental point of view, a most intriguing prediction of this class of non-commutative models is a small but different from zero probability for electrons to perform PEP violating atomic transitions ($\delta^2$), which depends on the energy scale of the observed transition. For both $\kappa$ and $\theta$ Poincar\'e, close to the non-commutativity scale $\Lambda$, the PEP violation probability turns to be of order one in the deformation parameter \cite{PRD_version}. For much smaller energies, the PEP violation probability is highly suppressed, accounting for the lack of evidence of PEP violation signals over decades of experimental efforts in this direction.

The experimental search for possible deviations from the PEP comprises several approaches, which depend on whether the superselection rule introduced by Messiah and Greenberg (MG) \cite{messiah} is fulfilled or not. MG states that, in a given closed system of identical fermions, the transition probability between two different symmetry states is zero.  PEP tests for electrons, which take into account MG, exploited: capture of $^{14}$C $\beta$ rays onto Pb atoms ($\delta^2 < 3 \cdot 10^{-2}$) \cite{Goldhaber1948}; pair production electrons captured on Ge ($\delta^2 < 1.4 \cdot 10^{-3}$) \cite{Elliott:2011cx}; PEP violating atomic transitions in conducting targets, performed by electrons introduced in the system by a direct current (best upper limit $\delta^2 < 8.6 \cdot 10^{-31}$) \cite{ramberg1990,curceanu2017,napolitano2022} or residing in the conduction band (best upper limit $\delta^2 < 1.53 \cdot 10^{-43}$) \cite{Elliott:2011cx,Piscicchia:2020kze}. MG does not apply to NCQG models; within this context strong bounds on the PEP violation probability, in atomic transitions, were set by the DAMA/LIBRA collaboration ($\delta^2 < 1.28 \cdot 10^{-47}$) \cite{bernabei2009}, searching for K-shell PEP violating transitions in iodine 
--- see also Refs.~\cite{ejiri,reines}. A similar analysis was performed by the MALBEK experiment ($\delta^2 < 2.92 \cdot 10^{-47}$) \cite{abgrall}, by constraining the rate of K$_\alpha$ PEP violating transitions in Germanium. 
We deploy a different strategy, without confining our analysis to the evaluation of a specific transition PEP violation probability. 
We consider PEP violating transition amplitudes, which we introduce in the next section, that enable a fine tuning of the $\theta$ tensor components. The spectral shape predicted for the whole complex of relevant transitions is tested against the data, constraining $\Lambda$, for the first time, far above the Planck scale for $\theta_{0i} \neq 0$.
Within a similar theoretical framework the DAMA/LIBRA limit on the PEP violating atomic transition probability \cite{bernabei2009} was analyzed in Ref.~\cite{Addazi:2018ioz}, and a lower limit on the non-commutativity scale $\Lambda$ was inferred as strong as $\Lambda > 5 \cdot 10^{16}$ GeV, corresponding to $\Lambda > 4 \cdot 10^{-3}$ Planck scales.\\

Resorting to different techniques, PEP violating nuclear transitions were also tested --- see e.g. Refs.~\cite{bernabei2009,bellini2010,suzuki1993}. The strongest bound ($\delta^2 < 7.4 \cdot 10^{-60}$) was obtained in Ref.~\cite{bellini2010}. The implications of these experimental findings for Planck scale deformed symmetries were investigated in Ref.~\cite{Addazi:2017bbg}, parametrizing the PEP-violation probability in terms of inverse powers of the non-commutativity scale. The analysis allowed to exclude a class of $\kappa$-Poincar\'e and $\theta$-Poincar\'e models in the hadronic sector. 
Nonetheless, within the context of NCQG models, tests of PEP in the hadronic and leptonic sectors need to be considered as independent. There is no a priori argument why fields of the standard model should propagate in the non-commutative space-time background being coupled to this latter in the same way. In string theory, for instance, non-commutativity emerges as a by-product of the constant expectation value of the B-field components, which in turn are coupled to strings' world-sheets with magnitudes that are not fixed a priori.
Constraints on $\delta^2$ were also inferred from astrophysical and cosmological arguments; the strongest bound ($\delta^2 < 2 \cdot 10^{-28}$) was obtained in Ref.~\cite{thoma1992}.\\

{\it Energy dependence of the PEP violation probability in NCQG models --- } The $\theta$-Poincar\'e model predicts (see Refs.~\cite{Addazi:2017bbg,11,12, PRD_version}) that PEP is violated with a suppression $\delta^2 =({E}/\Lambda)^{2}$, where ${E}\equiv {E}(E_1, E_2, \dots)$ is a combination of the characteristic energy scales of the transition processes under scrutiny (masses of the particles involved, their energies, the transitions energies etc.).
For a generic NCQG model deviations from the PEP in the commutation/anti-commutation relations can be parametrized \cite{Addazi:2017bbg} as $a_{i} a_{j}^{\dagger} -q(E) a_{j}^{\dagger} a_{i}=\delta_{ij}$, which resembles the quon algebra (see e.g. Refs.~\cite{Greenberg:1987aa,greenberg1991particles}), but has a Quantum Gravity induced energy dependence. While the q-model requires a hyper-fine tuning of the $q$ parameter, NCQG models encode $q(E)$, which is related to the PEP violation probability by $q(E)=-1+2\delta^{2}(E)$. 


For $\theta$-Poincar\'e models, taking into account two electrons of momenta $p_i^\mu=(E_i, \vec{p}_i )$ (with $i=1,2$), a phase $\phi_{\rm PEPV}$ can be introduced in order to parametrize the deformation of the standard transition probability $W_0$ into $W_\theta=W_0 \cdot \phi_{\rm PEPV}$. If we explicit the $\Lambda$ dependence in the $\theta$-tensor through the relation $\theta_{\mu \nu}= \tilde{\theta}_{\mu \nu}/\Lambda^2$, with $\tilde{\theta}_{\mu \nu}$ dimensionless, the energy scale dependence 
turns out to be: i) either
\begin{equation} \label{alzu}
\phi_{\rm PEPV} = \delta^ 2\simeq \frac{D}{2} \frac{E_N}{\Lambda} \frac{\Delta E}{\Lambda}\,,
\end{equation}
where $D=p_1^0 \tilde{\theta}_{0j} p_2^j + p_2^0 \tilde{\theta}_{0j} p_1^j$, the quantity $E_N\simeq m_N\simeq A \, m_{p}$ denotes nuclear energy and $\Delta E=E_2-E_1$ accounts for the atomic transition energy; ii) or 
\begin{equation} \label{cadu}
\phi_{\rm PEPV} = \delta^ 2 \simeq   \frac{C}{2} \frac{\bar{E}_1}{\Lambda} \frac{\bar{E}_2}{\Lambda}\,,
\end{equation}
where $\bar{E}_{1,2}$ are the energy levels occupied by the initial and the final electrons and $C= p_1^i \tilde{\theta}_{ij} p_2^j$. The former case, discussed in Eq.~\eqref{alzu}, encodes non-commutativity among space and time coordinates, namely $\theta_{0i}\neq 0$, while the latter case, in Eq.~\eqref{cadu}, corresponds to selecting $\theta_{0i}=0$, ensuring unitarity of the $\theta$-Poincar\'e models \cite{AlvarezGaume:2001ka,Gomis:2000zz}. In both cases the factors $D/2$ and $C/2$ can be approximated to unity.  \\ 


{\it The VIP-2 lead experiment ---}
The VIP-2 Lead experiment, operated at the Gran Sasso underground National Laboratory (LNGS) of INFN, realizes a dedicated high sensitivity test of the PEP violations for electrons, as observable signature of NCQG models.

The experimental setup is based on a high purity co-axial p-type germanium detector (HPGe), about 2 kg in mass. The detector is surrounded by a target, consisting of three 5 cm thick cylindrical sections of radio-pure Roman lead, for a total mass of about 22 kg (we refer to \cite{PRD_version,Piscicchia:2020kze,neder,heusser2006low} for a detailed description of the apparatus and the acquisition system). 
The strategy of the measurement is to search for PEP-violating $\mathrm{K}_{\alpha}$ and $\mathrm{K}_{\beta}$ transitions in the lead target, which occur when the $1s$ level is already occupied by two electrons. As a consequence of the additional electronic shielding, the energies of the transitions are shifted downwards, thus being distinguishable in a high precision spectroscopic measurement. In Table \ref{lines} the energies of the standard  $\mathrm{K}_{\alpha}$ and $\mathrm{K}_{\beta}$ transitions in Pb are reported, together with those calculated for the corresponding PEP-violating ones. The PEP violating K lines energies are obtained based on a multi configuration Dirac-Fock and General Matrix Elements numerical code \cite{indelicato}, see also Ref. \cite{Elliott:2011cx} where the $\mathrm{K}_{\alpha}$ lines are obtained with a similar technique.


\begin{table}[!h]
\caption{Calculated PEP-violating K${}_{\alpha}$ and K${}_{\beta}$ atomic transition energies in Pb
(column labeled forb.). As a reference, the allowed transition energies are also quoted (allow.). Energies are in keV.}
\label{lines}
\begin{center}
  \renewcommand\arraystretch{1.3}
\begin{tabular}{|c|c|c|}
\hline
\hline
  \textbf{Transitions in Pb}       & \textbf{allow. (keV)}  & \textbf{forb. (keV)}
\\ \hline
   1s - 2p${}_{3/2}$ K${}_{\alpha1}$  &  74.961 & 73.713 \\ \hline
   1s - 2p${}_{1/2}$ K${}_{\alpha2}$  &  72.798 & 71.652 \\ \hline
     1s - 3p${}_{3/2}$ K${}_{\beta1}$  &  84.939 & 83.856  \\ \hline
   1s - 4p${}_{1/2(3/2)}$ K${}_{\beta2}$  &  87.320 & 86.418   \\ \hline
     1s - 3p${}_{1/2}$ K${}_{\beta3}$  &  84.450 & 83.385  \\ \hline

\end{tabular}
\end{center}
\end{table}

All detector components were characterized and implemented into a validated Monte Carlo (MC) code (Ref. \cite{Boswell:2010mr}) based on the GEANT4 software library (Ref. \cite{Agostinelli:2002hh}), which allowed to estimate the detection efficiencies for X-rays emitted inside the Pb target.

The analyzed data sample corresponds to a total acquisition time $\Delta t\approx 6.1 \cdot 10^6 \mathrm{s} \approx 70 \, \mathrm{d}$, i.e. about twice the statistics used in Ref.~\cite{Piscicchia:2020kze}.   \\

{\it Data Analysis --- } We present the results of a Bayesian analysis, whose details are reported in the companion paper \cite{PRD_version}, aimed to extract the probability distribution function ({\it pdf}) of the expected number of photons emitted in PEP violating K$_\alpha$ and K$_\beta$ transitions. Comparison of the experimental upper bound on the expected value of signal counts $\bar{S}$, with the theoretically predicted value, 
provides a limit on the $\Lambda$ scale of the model. Let us notice that the algebra deformation preserves, at first order, the standard atomic transition probabilities, the violating transition probabilities being dumped by factors $\delta^2(E)$, hence transitions to the $1s$ level from levels higher than $4p$ will not be considered (see e.g. Ref. \cite{krause} for a comparison of the atomic transitions intensities in Pb). 

\begin{figure}[h]
\centering
\includegraphics[width=\columnwidth]{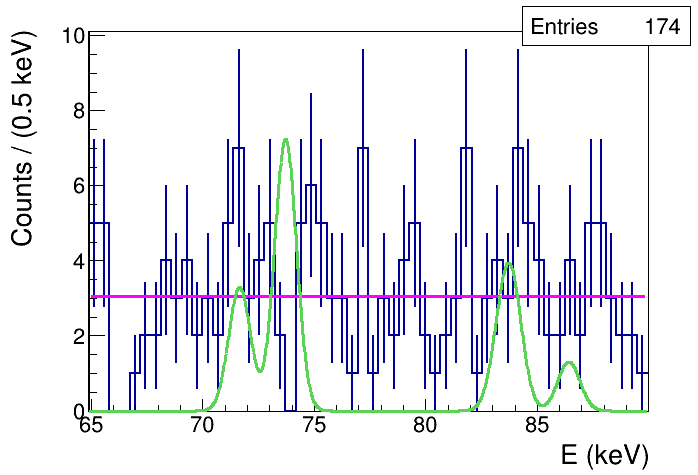}
\caption{
The measured X-ray spectrum, in the region of the K${}_{\alpha}$ and K${}_{\beta}$ standard and PEP-violating transitions in Pb, is shown in blue; the magenta line represents the fit of the background distribution. The green line corresponds to the shape of the expected signal distribution (with arbitrary normalization) for $\theta_{0i}\neq 0$.}
\label{spectrum}
\end{figure}

The measured energy spectrum is shown as a blue distribution in Figure \ref{spectrum}. Given the resolution of the detector ($\sigma$ better than 0.5 keV in $\Delta E = (65-90)$ keV) and a detailed characterization of the materials of the setup, the $\mathrm{K}_{\alpha}$ and $\mathrm{K}_{\beta}$ lead transitions are the only emission lines expected in the region of interest $\Delta E$. The target actively contributes to suppress background sources which  survive to the external passive shielding complex. Due to its extreme radio-purity, even the standard lead K complex can not be distinguished from a flat background, with average of $3$ counts/bin.  

The joint posterior $pdf$ of the expected number of total signal and background counts ($S$ and $B$) given the measured distribution - called \textit{data} - is:

\begin{equation}\nonumber
    P(S,B | data, \mathbf{p}) =
    \end{equation}
\begin{equation}
     =
     \frac{P(data | S,B,\mathbf{p}) \cdot f(\mathbf{p}) \cdot P_0 (S) \cdot P_0 (B) }{\int P(data | S,B,\mathbf{p}) \cdot f(\mathbf{p}) \cdot P_0 (S) \cdot P_0 (B) \ d^m\mathbf{p} \ dS \ dB} \, ,
    \label{f2}
\end{equation}
where $P_0$ denotes the prior distributions. To account for the uncertainties, introduced by both the measurement and the data analysis procedure, an average likelihood is considered, which is weighted over the joint $pdf$ of all the relevant experimental parameters $\mathbf{p}$. The likelihood is parametrised as:
\begin{equation}
 P(data | S,B,\mathbf{p}) = \prod_{i=1}^{N} \frac{\lambda_i (S,B,\mathbf{p})^{n_i} \cdot e^{-\lambda_i (S,B,\mathbf{p})}}{n_i !} \
\end{equation}
where $n_i$ are the measured bin contents. The number of events in the $i$-th bin fluctuates, according to a Poissonian distribution, around the mean value:

\begin{eqnarray}
    \lambda_i (S,B,\mathbf{p}) &=&  B \cdot \int_{\Delta E_i} f_B (E,\bm{\alpha}) \ dE + \\ \nonumber &+& S \cdot \int_{\Delta E_i} f_S (E,\bm{\sigma}) \ dE  
\ 
\end{eqnarray}
$\Delta E_i$ is the energy range corresponding to the $i$-th bin; 
$f_B (E,\bm{\alpha})$ and $f_S (E,\bm{\sigma})$ represent the shapes of the background and signal distributions normalised to unity over $\Delta E$.
Among the experimental uncertainties the only significant ones are those which characterize the shape of the background (parametrized by the vector $\bm{\alpha}$) and the resolutions ($\bm{\sigma}$) at the energies of the violating transitions (the resolutions are reported in Table II of Ref. \cite{PRD_version}). The rest of the experimental parameters are affected by relative uncertainties of the order of 1\% (or less), and are neglected; hence $\mathbf{p}=(\bm{\alpha}, \, \bm{\sigma})$.

The shapes for $f_S$ and $f_B$ are derived in the following Sections. \\

{\it Normalised signal shape ---} 
The rate of violating K$_{\alpha1}$ transitions predicted by the model, at the first order in the violation probability $\delta^2$, and weighted for the experimental detection efficiency, is derived in Ref. \cite{PRD_version}:

\begin{equation}\label{rate}
\Gamma_{K_{\alpha1}}= \frac{\delta^2(E_{K_{\alpha1}})}{\tau_{K_{\alpha1}}} \cdot \frac{BR_{K_{\alpha1}}}{BR_{K_{\alpha1}} + BR_{K_{\alpha2}}} \cdot 6 \cdot N_{atom} \cdot \epsilon(E_{K_{\alpha1}}). 
\end{equation}
In Eq. \eqref{rate} $E_K$ represents the proper combination of energy scales that enters Eqs.~\eqref{alzu}-\eqref{cadu}. $\tau_{K_{\alpha1}}$ is the lifetime of the PEP-allowed $2p_{3/2}  \rightarrow 1s$ transition (the lifetimes will be indicated with $\tau_K$ for the generic K transitions, their values from Ref. \cite{payne} are summarized in Table IV of Ref. \cite{PRD_version}, see also Ref. \cite{reines}). The branching fractions (which are given in Table \ref{roi}) allow to weight the relative intensities of the transitions which occur from levels with the same $(n,l)$ quantum numbers, but different $j$ (e.g. the $2p_{1/2}$ and the $2p_{3/2}$). $N_{atom}$ accounts for the total number of atoms in the lead sample. The efficiencies for the detection of photons emitted in the target, at the energies corresponding to the violating transition lines, are listed in  Table \ref{roi}. The expected number of PEP violating $K_{\alpha1}$ events measured in $\Delta t$ is then

\begin{table}
\caption{The table summarizes the values of the branching ratios of the considered atomic transitions and the detection efficiencies at the energies corresponding to the
K${}_{\alpha}$ and K${}_{\beta}$ PEP-violating transitions.
}
\label{roi}
\begin{center}
   \renewcommand\arraystretch{1.3}
\begin{tabular}{|c|c|c|}
\hline
\hline
   \textbf{PEP-viol. trans.}  & $BR$ & $\epsilon$
   \\ \hline
    K${}_{\alpha1}$
     & 0.462 $\pm$ 0.009 & $(5.39\pm 0.11)\cdot 10^{-5}$ \\ \hline
    K${}_{\alpha2}$
     & 0.277 $\pm$ 0.006 & $(4.43^{+0.10}_{-0.09})\cdot 10^{-5}$  \\ \hline
       K${}_{\beta1}$
     & 0.1070 $\pm 0.0022$ & $(11.89 \pm 0.24)\cdot 10^{-5}$ \\ \hline
    K${}_{\beta2}$
     & 0.0390 $\pm$ 0.0008 & $(14.05^{+0.29}_{-0.28})\cdot 10^{-5}$  \\
\hline
K${}_{\beta3}$
     & 0.0559 $\pm$ 0.0011 & $(11.51^{+0.24}_{-0.23})\cdot 10^{-5}$  \\
\hline
\end{tabular}
\end{center}
\end{table}

\begin{equation}\label{expnum}
\mu_{K_{\alpha1}} = \Gamma_{K_{\alpha1}} \cdot \Delta t\,.
\end{equation}
The expected number of counts for any PEP violating K transition is obtained by analogy with Eq. \eqref{expnum}. 

The probability of two (or more) steps processes, involving transitions from higher levels to the $n p$ one ($n=2,3,4$), followed by the violating K transition, 
scales as the product of the corresponding $\delta^2$ terms and is neglected at the first order. The same argument also holds for subsequent violating transitions from the same atomic shell $n p$ ($n=2,3,4$) to $1s$.

$f_S (E)$ is then given by the sum of Gaussian distributions, whose mean values ($E_K$) are the energies of the PEP violating transitions in Pb, and the widths ($\sigma_K$) are the resolutions at the corresponding energies. The amplitudes are weighted by the rates $\Gamma_K$ of the corresponding transitions (see Eq. \eqref{rate}): 

\begin{equation}\label{fs}
    f_S (E) = \frac{1}{N} \cdot \sum_{K=1}^{N_K} \Gamma_K \frac{1}{\sqrt{2 \pi \sigma_K^2}}  \cdot e^{-\frac{(E - E_K)^2}{2 \sigma_K^2}}.
\end{equation}
It is important to note that the $\Gamma_K$ term in Eq. \eqref{fs} entails a dependence on the 
$\theta_{0i}$ choice (through the proper energy dependence term) which is contained in $\delta^2$ see Eqs.~\eqref{alzu}-\eqref{cadu}. For this reason two independent analyses are performed for the two $\theta_{0i}$ cases, by following the same procedure, in order to set constraints on the $\Lambda$ scale of the corresponding specific model. $f_S$ does not itself depend on $\Lambda$, since the dependence is re-absorbed by the normalisation, which is given by

\begin{equation}\label{norm}
    \int_{\Delta E} f_S (E) dE = 1 \Rightarrow N = \sum_{K=1}^{N_K}
    \Gamma_K.
\end{equation} 
In Eqs. \eqref{fs} and \eqref{norm} the sum extends over the number $N_K$ of the PEP violating transitions listed in Table \ref{lines}.

As an example, the shape of the expected signal distribution, for $\theta_{0i}\neq 0$, is shown with arbitrary normalization as a green line in Figure \ref{spectrum}.\\

{\it Normalised background shape ---} 
In order to determine the shape of the background a maximum log-likelihood fit of the measured spectrum is performed, excluding 3$\sigma_K$ intervals centered on the mean energies $E_K$ of each violating  transition.
The best fit yields a flat background amounting to  $L(E) = \alpha = ( 3.05   \pm 0.29) \ \mathrm{counts/(0.5\,\mathrm{keV})}$  (the errors account for both statistical and systematic uncertainties),
corresponding to $f_B (E) = {L(E)}/{\int_{\Delta E} L(E) \ dE}.$\\

{\it Prior distributions --- } $P_0(B)$ is taken to be Gaussian for positive values of $B$ and zero otherwise. The expectation value is given by $B_0 = \int_{\Delta E} L(E) \ dE $. For comparison a Poissonian prior was also tested for $B$. The upper limit on $\bar{S}$ is not affected by this choice, within the experimental uncertainty.

Given the a priori ignorance, a uniform prior is assumed for $S$, in the range ($0 \div S_{max}$). $S_{max}$ is the maximum number of expected X-ray counts, from PEP violating transitions in Pb, according to the best independent experimental limit (Ref. \cite{Elliott:2011cx}).
$S_{max}$ is then given by Eq. 3 of Ref. \cite{Elliott:2011cx}, evaluated by substituting the parameters which characterise our experimental apparatus (see Tables \ref{roi} and \ref{parameters}).
We obtain $S_{max} \approx 1433$ and $P_0(S) = 1/S_{max} \ \left[ \Theta(S) - \Theta(S - S_{max}) \right]$,
where $\Theta$ is the Heaviside function. \\

\begin{table}[!h]
\caption{Values of the parameters which characterise the Roman lead target, from left to right: free electron density, volume, mass and number of free electrons in the conduction band.}
\label{parameters}
\begin{center}
 \renewcommand\arraystretch{1.3}
\begin{tabular}{|c|c|c|c|c|c|}
\hline
\hline
$n_e$(m$^{-3}$)  & $V$(cm$^3$) & $M$(g) & $N_\mathrm{free}$ \\ \hline
$1.33 \cdot 10^{29}$ & $2.17 \cdot 10^{3}$ & 22300 & $2.89 \cdot 10^{26}$ \\
\hline
\end{tabular}
\end{center}
\end{table}

\begin{figure}
    \centering
    \includegraphics[width=\columnwidth]{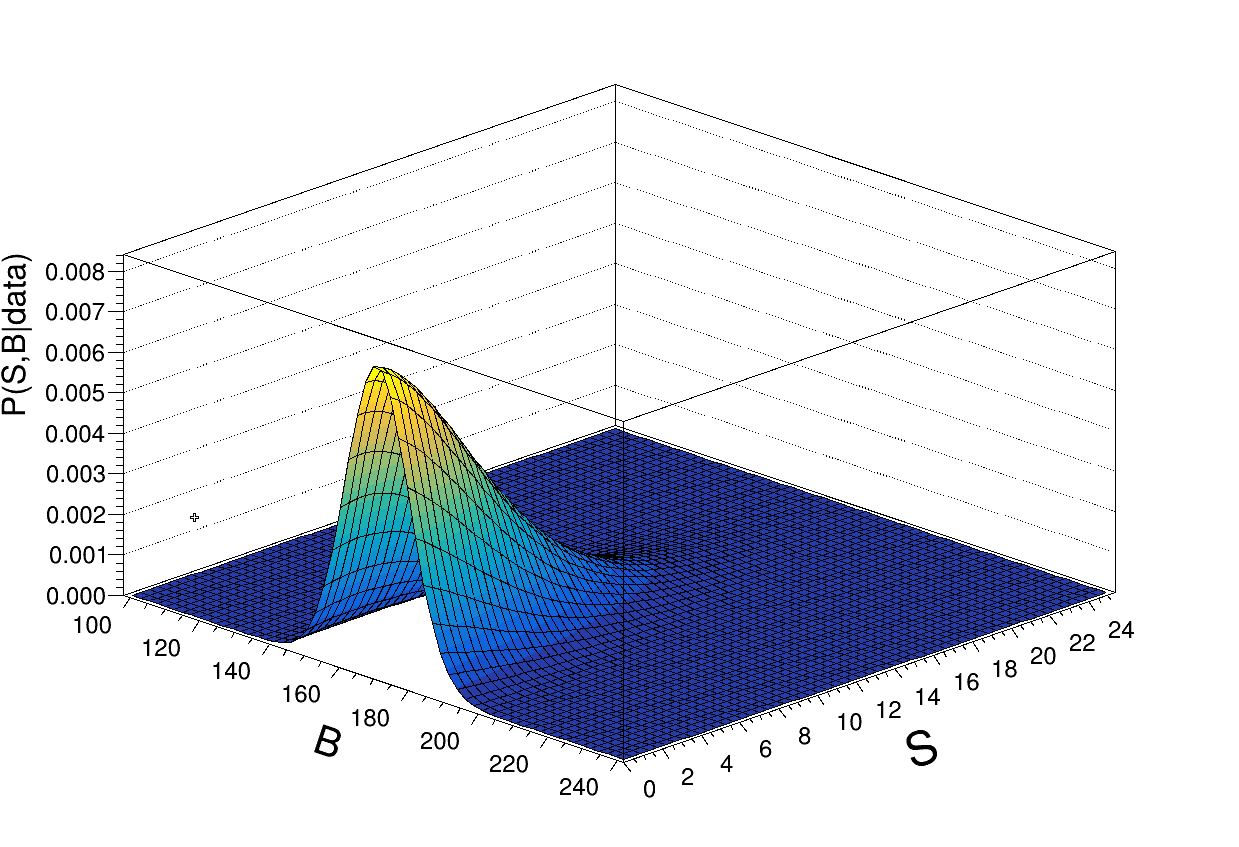}
    \caption{Joint $pdf$ $P(S,B|data)$ of the expected number of total signal and background counts corresponding to  $\theta_{0i} \neq 0$.
    .}
    \label{fig:joint}
\end{figure}

{\it Lower limits on the non-commutativity scale $\Lambda$ --- } The upper limits $\bar{S}$ are calculated, for each choice of $\theta_{0i}$, by solving the following integral equation for the cumulative distribution $\tilde{P}(\bar{S})$:

\begin{equation}\label{cumulative}
        \tilde{P}(\bar{S}) = \int_0^{\bar{S}} P(S|data) \ dS = \Pi,
\end{equation}
The posterior $pdf$ and the cumulative distribution are calculated by means of a dedicated algorithm. Numerical integrations are performed following Monte Carlo techniques, a detailed description is provided in the appendix of Ref.~\cite{PRD_version}. As an example the joint $pdf$ $P(S,B|data)$ is shown in Figure \ref{fig:joint} for $\theta_{0i} \neq 0$. We obtain $\bar{S}<13.2990$ and $\bar{S}<18.1515$, with a  probability $\Pi = 0.9$, respectively for $\theta_{0i} = 0$ and $\theta_{0i} \neq 0$. The results are affected by a relative numerical error of $\sim 2 \cdot 10^{-5}$.

A direct comparison of the total predicted violating transitions expected number $\mu$ and the corresponding upper bound $\bar{S}$, namely

\begin{equation}
\mu = \sum_{K=1}^{N_K} \mu_K = 
\frac{\aleph}{\Lambda^2} < \bar{S} \Rightarrow \Lambda > \left(  \frac{\aleph}{\bar{S}} \right)^{1/2},
\end{equation}
provides the following lower limits on the non-commutativity scale: 

\begin{itemize}
    \item $\Lambda > 6.9\cdot 10^{-2}$ Planck scales $\qquad$  for $\theta_{0i} = 0$
    \item $\Lambda > 2.6\cdot 10^{2}$ Planck scales $\qquad$ for $\theta_{0i} \neq 0$
\end{itemize}
corresponding to a probability $\Pi = 0.9$.\\

{\it Discussion and conclusions ---}
The analysis of the total data set collected by the VIP-2 Lead collaboration is presented. The experiment is designed for a high sensitivity search of Pauli Exclusion Principle (PEP) violations in atomic transitions. 
Upper limits are set on the expected signal of PEP violating K$_\alpha$ and K$_\beta$ transitions, generated in a high radio-purity Roman lead target, by means of a Bayesian comparison of the measured spectrum with the violating K complex shape predicted by the $\theta$-Poincar\'e Non Commutative Quantum Gravity (NCQG) model. 

The analysis yields stringent bounds on the non-commutativity energy scale, which exclude $\theta$-Poincar\'e up to 2.6$\cdot 10^2$ Planck scales when the ``electric like" components of the $\theta_{\mu \nu}$ tensor are different from zero, and up to 6.9$\cdot 10^{-2}$ Planck scales if they vanish, thus providing the strongest (atomic-transitions) experimental test of the model.

The most intriguing theoretical feature --- see e.g. Eqs. \eqref{alzu} and \eqref{cadu} --- consists in a strong dependence of the predicted departure from PEP on the energy scales involved in the analyzed process. A systematic study of data from ongoing   \cite{bernabei2009,abgrall} and forthcoming experiments, 
in analogy with the analyses of Refs.~\cite{Addazi:2017bbg,Addazi:2018ioz} while focusing on signatures of atomic PEP violation, would substantially supplement our conclusions.
The VIP collaboration is presently implementing an upgraded experimental setup, based on cutting-edge Ge detectors, aiming to probe $\theta$-Poincar\'e beyond the Planck scale, independently of the particular choice of the $\theta_{\mu \nu}$ electric like components. 
NCQG models, in a large number of their popular implementations, are being tested and eventually ruled-out. In this sense, contrary to naive expectations, NCQG is not only a theoretical attractive mathematical idea, but also source of a rich phenomenology, which can be tested in high sensitivity X-ray spectroscopic measurements.

\section*{Acknowledgments}
\noindent
This publication was made possible through the support of Grant 62099 from the John Templeton Foundation. The opinions expressed in this publication are those of the authors and do not necessarily reflect the views of the John Templeton Foundation.
We acknowledge support from the Foundational Questions Institute and Fetzer Franklin Fund, a donor advised fund of Silicon Valley Community Foundation (Grants No. FQXi-RFP-CPW-2008 and FQXi-MGB-2011), and from the H2020 FET TEQ (Grant No. 766900). 
We thanks: the INFN Institute, for supporting the research presented in this article and, in particular, the Gran Sasso underground laboratory of INFN, INFN-LNGS, and its Director, Ezio Previtali, the LNGS staff, and the Low Radioactivity laboratory for the experimental activities dedicated to the search for spontaneous radiation; the Austrian Science Foundation (FWF), which supports the VIP2 project with the grants P25529-N20, project P 30635-N36 and W1252-N27 (doctoral college particles and interactions). 
K.P. acknowledges support from the Centro Ricerche Enrico Fermi - Museo Storico della Fisica e Centro Studi e Ricerche “Enrico Fermi” (Open Problems in Quantum Mechanics project). 
A.A. work is supported by the Talent Scientific Research Program of College of Physics, Sichuan University, Grant No.1082204112427 \& the Fostering Program in Disciplines Possessing Novel Features for Natural Science of Sichuan University,  Grant No. 2020SCUNL209 \& 1000 Talent program of Sichuan province 2021. 
AM wishes to acknowledge support by the Shanghai Municipality, through the grant No. KBH1512299, by Fudan University, through the grant No. JJH1512105, the Natural Science Foundation of China, through the grant No. 11875113, and by the Department of Physics at Fudan University, through the grant No. IDH1512092/001. 
A.A and A.M. would like to thank Rita Bernabei and Pierluigi Belli for useful discussions on this subject.

\end{document}